\documentstyle[twocolumn,epsf,aps]{revtex}
\newcommand{\postscript}[2] {\setlength{\epsfxsize}{#2\hsize}
\centerline{\epsfbox{#1}}}
\begin{document}
\twocolumn[\hsize\textwidth\columnwidth\hsize\csname 
@twocolumnfalse\endcsname
\title{Many-body system with a four-parameter family of 
point interactions in one dimension}
\author{F A B Coutinho$^a$, Y Nogami$^{b,c}$ and 
Lauro Tomio$^c $}
\address{
$^a$ Faculdade de Medicina, Universidade de S\~ao Paulo, 
01246-903, S\~ao Paulo, Brazil \\
$^b$ Department of Physics and Astronomy, McMaster University,\\  
Hamilton, Ontario, Canada L8S 4M1\\
$^c$ Instituto de F\'{\i}sica Te\'orica, Universidade Estadual 
Paulista, 01405-900, S\~ao Paulo, Brazil
}
\date{\today}
\maketitle
\begin{abstract}
We consider a four-parameter family of point interactions in 
one dimension. This family is a generalization of the usual 
$\delta$-function potential. We examine a system consisting of 
many particles of equal masses that are interacting pairwise 
through such a generalized point interaction. We follow McGuire 
who obtained exact solutions for the system when the interaction 
is the $\delta$-function potential. We find exact bound 
states with the four-parameter family. For the scattering 
problem, however, we have not been so successful. This is 
because, as we point out, the condition of no diffraction that 
is crucial in McGuire's method is not satisfied except when 
the four-parameter family is essentially reduced to the 
$\delta$-function potential.
\end{abstract}\vskip 0.2cm
]

\section{Introduction}
One of the exactly solvable many-body models in quantum 
mechanics is a system of particles of equal masses in one 
dimension, interacting through a $\delta$-function potential 
\cite{1,2}. The $\delta$-function potential is a special 
case of a large family of point interactions in one dimension. 
The family of point interactions represents all possible 
self-adjoint extensions of the kinetic energy operator 
\cite{3,4,5,6,7}. There are two types of point interactions, 
penetrable and impenetrable. An impenetrable point 
interaction separates the space into two disjoint 
half-spaces. In this paper we focus on the penetrable type 
which we think is more interesting in physics than the 
impenetrable type. The point interactions of the penetrable 
type can be specified in terms of four real parameters.

It has recently been pointed out that, for a one-body problem 
in which a particle interacts with a given point interaction, 
three parameters are actually sufficient \cite{8}. The fourth 
parameter, denoted with $\theta$ in the following, is 
redundant. This is in the sense that, although the wave 
function of the particle depends on $\theta$, observable 
quantities like the transmission and reflection probabilities, 
the energy eigenvalue and the probablity density of a bound 
state are all independent of $\theta$. In many-body problems, 
however, $\theta$ may have subtle implications in relation to 
the symmetry of the wave function.

The purpose of this paper is to examine the three-body and 
many-body problems in one dimension when the particles 
involved have equal masses and are interacting through one 
of the penetrable point interactions of the four-parameter 
family. We assume that the interaction is common to all pairs. 
It is understood that the particles have no spin. We treat 
the particles as distinguishable ones without imposing any 
symmetry requirements on the many-body wave function. In 
certain situations the wave function becomes symmetric or 
antisymmetric with respect to interchanges of the particles.
In such cases the wave function can be interpreted as that
for bosons or fermions.

This work is an extension of part of McGuire's pioneering 
work in which the same problems were exactly solved with the 
$\delta$-function potential \cite{1}. For the bound state 
we find that McGuire's solutions can easily be extended to 
accommodate the four-parameter family. The scattering problem 
is much harder. We find that the condition of no diffraction 
that is crucial in McGuire's method for the scattering problem 
is satisfied only if the four parameters are restricted in a 
certain manner. For the general four-parameter family of point 
interactions, the scattering problem requires more 
sophisticated approach, which is beyond the scope of the 
present paper. 
\footnote{
After submitting an earlier version of this paper for 
publication in this journal, one of the referees kindly 
brought a preprint by Albeverio et al (ADF) \cite{9} to our 
attention. ADF independently addressed essentially the same 
problems. For the point interactions of the penetrable type, 
our scattering solutions agree with theirs. In the bound 
state problem, however, our solutions are more general than 
theirs. We will comment on ADF's work towards the end of 
this paper.}

In section 2 we summarize relevant aspects of the one-body 
and two-body problems with the point interactions. In section 
3 we determine the three-body and many-body bound states. 
Section 4 is devoted to the scattering states. Summary and 
discussion are given in section 5. There are two appendices. 
In appendix 1 we discuss the symmetry aspect the many-body 
wave function in relation to parameter $\theta$. In appendix 2 
we summarize the results for the $\delta$-function potential 
so that our results can easily be compared with those of 
McGuire.

\section{Point interactions}
Let us start with a one-body problem; a particle interacting 
with a given point interaction at $x=0$. A point interaction 
is such that it is zero everywhere except at $x=0$. The point 
interaction can be interpreted in terms of self-adjoint 
extension of the nonrelativistic kinetic energy operator 
$-(\hbar^2/2m)d^2/dx^2$ where m is the mass of the particle 
concerned. In the following we use units in which $\hbar = 1$.
Note that, as we stated in section 1, we do not consider the 
impenetrable type of point interactions that disconnect the 
half-spaces of $x>0$ and $x<0$.
 
It is known that there are four-parameter family of 
penetrable point interactions \cite{3,4,5,6,7}. They can be 
expressed in terms of the boundary condition on the wave 
function at $x=0$. The boundary condition can be written as
\begin{equation}
\left( \begin{array}{c}\psi^{\prime}(+0) \\ 
2m\psi(+0) \end{array} \right)
= U \left( \begin{array}{c}\psi^{\prime}(-0) \\ 
2m\psi(-0) 
\end{array} \right),
\label{1}
\end{equation}
\begin{equation}
U= e^{i\theta} \left( \begin{array}{cc}\alpha & \beta \\
                            \delta & \gamma 
\end{array} \right), 
\hspace{0.3 cm} \alpha \gamma -\beta \delta = 1\, , 
\label{2}
\end{equation}
where $\psi ' (x) = d\psi(x)/dx$ and $\alpha, \beta ,\gamma,  
\delta$ and $\theta$ are real dimensionless constants. Among 
$\alpha, \beta, \gamma$ and $\delta$, three are independent. 
Thus we have a four-parameter family of point interactions. 

It would be useful to relate the parameters specified above 
to another set of real parameters $a$, $b$, $c$ and $d$
that have appeared in the literature \cite{6,9,10}. 
In these references units are chosen such that $2m=1$.
In terms of these parameters $U$ can be expressed as
\begin{equation}
U= e^{i\theta}\left( \begin{array}{cc} d & c \\ b & a
\end{array} \right), \hspace{0.3 cm} ad - bc = 1 \, . 
\label{3}
\end{equation}
The $\theta$ is common between (\ref{2}) and (\ref{3}).
The other parameters are related by $a=\gamma$, $b=\delta$,
$c=\beta$ and $d=\alpha$. In \cite{6}, $\omega =e^{i\theta}$
was actually used. In the present paper we use the notation 
of (\ref{1}) and (\ref{2}) throughout.

For a point interaction that is expressed as above, we can 
work out all physics problems such as those of transmission, 
reflection and bound state. It turns out that, although the 
wave function obviously depends on $\theta$ through the phase 
factor $e^{i\theta}$, all physically observable quantities
such as various probabilities, matrix elements, energy 
eigenvalues are independent of $\theta$. In this sense 
$\theta$ is a redundant parameter. Two point interactions 
that differ only through the choice of the value of $\theta$ 
are physically equivalent. If $e^{i\theta}$ is complex, it 
may look as if time-reversal invariance is violated. This 
is actually not the case \cite{8}. For a one-body system, it
is therefore sufficient to take the three-parameter family 
without $\theta$, that is, by keeping $\theta$ fixed to an 
arbitrary value. 

In the many-body case that we study in the following 
sections, all physical quantities such as the energy 
eigenvalues and various matrix elements will be independent 
of $\theta$. In this sense $\theta$ is again redundant. 
The wave function, however, depends on the choice of 
$\theta$. This will have relevance regarding the
symmetry, if any, of the wave function. 
If $e^{i\theta}$ is complex, the wave function does not 
seem to have any interesting symmetry. 
If $e^{i\theta} = \pm 1$ and $\alpha = \gamma$,
we will see that the wave function exhibits simple 
symmetries. In the following we retain $\theta$ but 
occasionally we focus on the cases of $e^{i\theta}=\pm 1$.

Suppose that the interaction is invariant under space 
reflection $x \to -x$. This means that the boundary condition
is invariant under $\psi (\pm 0) \to \psi (\mp 0)$ and 
$\psi ^\prime (\pm 0) \to -\psi ^\prime (\mp 0)$. This holds 
if and only if $\alpha = \gamma$ and $e^{i\theta} = \pm 1$. 
Let us mention two special cases. For the familiar 
$\delta$-function potential $V(x) = g\delta (x)$ we obtain 
\begin{equation}
\alpha =-1\, ,\;\;\beta =-g\, ,\;\;\gamma =-1\, ,
\;\;\delta =0\,,\;\; e^{i\theta} = -1\,.
\label{4}
\end{equation}
On the other hand, the so-called $\delta ^\prime $ 
interaction\cite{3,4,6,7} is defined by the boundary 
condition with 
\begin{equation}
\alpha =-1\, ,\;\;\beta =0,\, \;\;\gamma =-1\, ,
\;\;\delta =-c \, ,\;\; e^{i\theta} = -1\, .
\label{5}
\end{equation}
where $c$ is a constant. This implies that, while $\psi '(x)$ 
is continuous at $x=0$, $\psi (x)$ is discontinuous. The 
$\delta ^\prime $ interaction so defined is invariant under 
$x \to -x$ (because $\alpha = \gamma$ and $e^{i\theta}$ is
real). It was already emphasized in \cite{7} that the 
$\delta ^\prime $ interaction has little resemblance to what 
the name may suggest (i.e. $d \delta (x)/dx$).

In defining the above two special interactions we have 
chosen $\theta$ such that $e^{i\theta} = -1$. This is to 
conform to the notation that was used earlier in \cite{4,5,7}. 
If we opt for $e^{i\theta} = 1$, then the signs of
the other parameters are simply reversed. Actually this
latter choice seems more convenient. To avoid any 
unnecessary confusion, however, we will adhere to (\ref{4})
and (\ref{5}). For the choice of $e^{i\theta} = 1$, see 
appendix 1 also. 

If the interaction is effectively attractive, there can be 
one or two bound states \cite{6,7}. The wave function of 
a bound state is of the form of
\begin{equation}
\psi _{\pm}(x) = C_{\pm} e^{-\kappa |x|}\, ,
\label{6}
\end{equation}
where $\kappa >0$ and the suffix $\pm$ refers to the sign of 
$x$. The energy of the bound state is given by 
\begin{equation}
E = -\frac{\kappa ^2}{2m}\, .
\label{7}
\end{equation}
The boundary condition (\ref{1}) requires that
\begin{equation}
\delta \kappa^2 + 2(\alpha +\gamma)\kappa m 
+ 4\beta m^2 =0\, ,
\label{8}
\end{equation}
which leads to
\begin{equation}
\frac{\kappa}{2m} = \left\{ \begin{array}{l}
{\displaystyle 
\frac{1}{2\delta}}[-(\alpha +\gamma) 
\pm \sqrt{(\alpha -\gamma)^2 +4\, }\, ] 
\hspace{0.4 cm} {\rm if} \,\, \delta \neq 0\, , \\
{\displaystyle 
-\frac {\beta}{\alpha +\gamma} \hspace{0.4 cm} }
{\rm if} \,\, \delta =0 \;\; {\rm and} \;\; \alpha + \gamma 
\neq 0\, . \end{array} \right.
\label{9}
\end{equation}
If equation (\ref{8}) for $\kappa$ has a positive real root, 
there is a bound state of energy $E=-\kappa ^2 /(2m)$. 
For the $\delta$-\-function potential of (\ref{4}) with $g<0$, 
we obtain $\kappa = -gm$. For the $\delta '$ interaction of 
(\ref{5}) with $c<0$, we find $\kappa = -4m/c$. 

In general there can be two bound states. For example, if 
$\delta \neq 0$ and $\alpha = \gamma$, (\ref{9}) can be 
reduced to
\begin{equation}
\frac{\kappa}{2m} = \frac{-\alpha \pm 1}{\delta}.
\label{10}
\end{equation}
If $\delta >0$ and $\alpha <-1$, there are two positive 
roots for $\kappa$. The $+$ of the double sign corresponds 
to the ground state. If $\delta >0$ and $\alpha =-1$, this 
is a $\delta '$ interaction of (\ref{5}) with $c<0$. In this 
case there is only one bound state. Figure 1 shows the areas 
in the $\alpha$-$\gamma$ plane in which there are zero, one 
and two bound states.

\vskip -0.4cm
\begin{figure}
\postscript{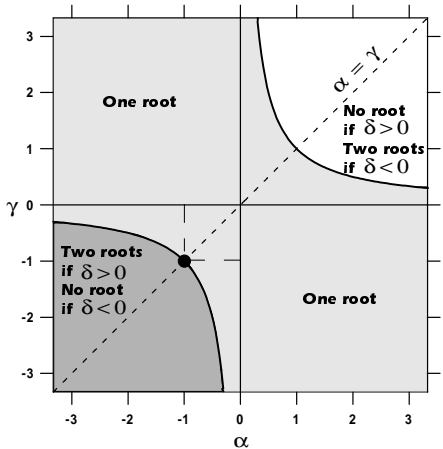}{0.9}    
\vskip -1cm
\caption{The regions in which there are 0, 1 or 2 real
positive roots of (\ref{8}) and hence 0, 1 and 2 bound
states.}
\label{fig1}
\end{figure}

The $\psi(x)$ and $\psi ' (x)$ are generally discontinuous 
at $x=0$ \cite{6,7}. We obtain the ratio
\begin{equation}
\eta \equiv \frac {\psi (+0)}{\psi (-0)} = \frac {C_+}{C_-} 
= -e^{i\theta}\left(\alpha + \frac{2\beta m}{\kappa}\right) 
= e^{i\theta}\left( \gamma + \frac{\delta \kappa}{2m}\right) .
\label{11}
\end{equation}
It can be shown that $|\eta|=|\psi (+0)/\psi (-0)|=1$ if and 
only if $\alpha=\gamma$. 

Let us consider the case in which there are two bound states. 
It is understood that $\delta \neq 0$. In this case the 
ratio of (\ref{11}) can be reduced to
\begin{equation}
\eta = -\frac{1}{2}e^{i\theta}\left[\alpha -\gamma 
\mp\sqrt{(\alpha -\gamma) ^2+ 4\,}\, \right]\, .
\label{12}
\end{equation}
If we distinguish the $C_{\pm}$'s of the two bound states
by adding superscripts $(\pm)$ that correspond to the $\pm$ 
of (\ref{9}) and (\ref{10}) or the $\mp$ of (\ref{12}), we 
obtain
\begin{equation}
C^{*(+)}_+ C^{(-)}_+ + C^{*(+)}_- C^{(-)}_- = 0\, .
\label{13}
\end{equation}
By using this relation one can show that the wave functions 
of the two bound states are orthogonal to each other.

Let us look into the special case of $\alpha = \gamma$ and 
$\delta \neq 0$. In this case we obtain
\begin{equation}
\eta = \pm e^{i\theta}\, ,
\label{14}
\end{equation}
where the double sign corresponds to those of (\ref{9}) and
(\ref{10}). Suppose we choose $\theta$ as $e^{i\theta}=1$. 
If there are two bound states, as we discussed below 
(\ref{10}), the parity is even for the ground state and odd 
for the excited state. If we choose $\theta$ as 
$e^{i\theta}=-1$, the parity of each of the states is 
reversed. We will discuss this aspect more in appendix 1. 

In addition to the bound state problem, the problem of 
transmission and reflection can also be worked out 
\cite{6,7}. We will quote the transmission and reflection 
coefficients in section 4 where we will examine the 
three-body scattering problem. 

Before proceeding to the three-body and many-body systems 
that we examine in the next section, it would be useful to 
briefly examine the two-body system. Let us introduce the 
variable $x$ defined by
\begin{equation}
x= \frac{1}{\sqrt 2}(x_1 -x_2)\, .
\label{15}
\end{equation}
This $x$ differs from the usual relative coordinate by a 
factor of $\sqrt 2$. We use this because this is one of the
Jacobi coordinates that are commonly used in the three-body 
problem. In this way we are treating the two-body system as 
a subsystem of a three-body system. The centre-of-mass 
coordinate can be separated as usual and the Schr\"{o}dinger 
equation for the system can be reduced to 
\begin{equation}
\left[- \frac{1}{2m}\frac{d^2}{dx ^2} + V(x) \right] \psi (x) 
= E \psi (x)\, ,
\label{16}
\end{equation}
where $V(x)$ stands for the point interaction that is defined 
by the boundary condition (\ref{1}) together with (\ref{2}). 
The energy $E$ does not contain the part that is due to the 
centre-of-mass motion. 

The interaction $V(x)$ of (\ref{16}) can be a source of 
confusion. Recall that the inter-particle distance is 
$x_1 -x_2 = \sqrt{2} x$. To make the problem clear, let us 
consider the usual $\delta$-\-function potential. Suppose 
that we start with the two-body interaction $V(x_1 -x_2) = 
g_0 \delta (x_1 -x_2)$ and change the variable to $x$, 
we obtain $V(x_1 -x_2) = g_0 \delta(\sqrt{2} x) 
= (g_0 /\sqrt{2})\delta(x)$. In this interpretation, the 
interaction of (\ref{16}) should be $V(\sqrt{2} x)$ rather 
than $V(x)$. In this connection, see appendix 2. For a 
generalized interaction, the boundary condition of 
(\ref{1}) has to be appropriately scaled.

In this paper we take a different interpretation. Instead 
of starting with $V(x_1 -x_2)$ and scaling the interaction 
and the boundary condition as we described above, 
let us take the $V(x)$ of (\ref{16}) as the one defined by 
(\ref{1}) and (\ref{2}) with the understanding that $x$ is 
the variable defined by (\ref{15}). After all we take this 
as a matter of definition of the two-body interaction. The 
main issue that we want to focus on is, with the two-body 
interaction so defined, how the three-body problem can be 
solved. A great advantage of this definition of $V(x)$ of 
(\ref{16}) is that all the formulae that we have obtained
for the one-body problem can be used for the two-body
and many-body problems with the understanding that $x$ is 
the one defined by (\ref{15}). For a bound state of the
two-particle  system, the wave function is given by 
(\ref{6}) and its energy by (\ref{7}), and so on.

\section{Three-body and many-body bound states}

We consider a system of many particles of equal masses,
interacting through a point interaction that is represented 
by boundary condition (\ref {1}). We begin with the 
three-body problem. Let the coordinates of the three 
particles be $x_1$, $x_2$ and $x_3$, and introduce the 
Jacobi coordinates $x$, $y$ and $z$ by
\begin{eqnarray}
&&x=\frac{1}{\sqrt 2}(x_1 -x_2)\, , \;\;\; 
y=\sqrt {\frac {2}{3}} 
\left( \frac {x_1 +x_2}{2} - x_3 \right) , \nonumber\\
&&z= \frac {1}{\sqrt 3}(x_1 +x_2 +x_3)\, .
\label{17} \\
&&\frac {x - \sqrt{3} y}{2} 
= \frac{-(x_2 -x_3)}{\sqrt 2}\, , \;
\frac {x + \sqrt{3} y}{2} = \frac{-(x_3 -x_1)}
{\sqrt 2}\, .
\label{18}
\end{eqnarray}
The Schr\"{o}dinger equation for the three-body system reads
\begin{eqnarray}
&&\left[- \frac{1}{2m} \left( \frac{\partial ^2}{\partial x^2} 
+ \frac{\partial ^2}{\partial y^2} \right) 
+ V(x) + V\left(-\frac{x - \sqrt{3} y}{2} \right)
\right.\nonumber\\ &&\left.
+ V\left(-\frac{x + \sqrt{3} y}{2} \right)
\right] \psi (x,y) = E\psi (x,y)\, .
\label{19}
\end{eqnarray}
It is understood that the coordinate $z$ (that is essentially 
the centre-of mass coordinate) has been separated already.
It is also understood that the two-body interactions like
$V(x)$ are interpreted as in section 2, below (\ref{14}).

Let us consider six regions that are specified in terms of 
the signs of $x_{12}$, $x_{23}$ and $x_{31}$, where 
$x_{ij}=(x_i -x_j)/\sqrt{2}$. \ We designate the regions with 
1, 2, $\cdots$ , 6, or with $(++-)$, $(-+-)$, $\cdots$ , 
$(+--)$; see figure 2. For example, in region 1, $x_{12}>0$, 
$x_{23}>0$ and $x_{31}<0$.  Note that $(+++)$ and $(---)$ are 
not possible. Our six regions 1, 2, $\cdots$ , 6 correspond
to McGuire's regions II, I, III, V, VI and IV, in this order.

Let us assume that the interaction is effectively attractive 
and there is a bound state. In each of the six regions, the 
Schr\"{o}dinger equation for the three-particle system is 
satisfied by
\begin{eqnarray} 
\phi (x,y) = e^{-\kappa \left(|x|
+ \frac{|x + \sqrt{3} y|}{2}
+ \frac{|x - \sqrt{3} y|}{2} \right)} ,
\label{20}
\end{eqnarray}

\vskip -0.5cm
\begin{figure}
\postscript{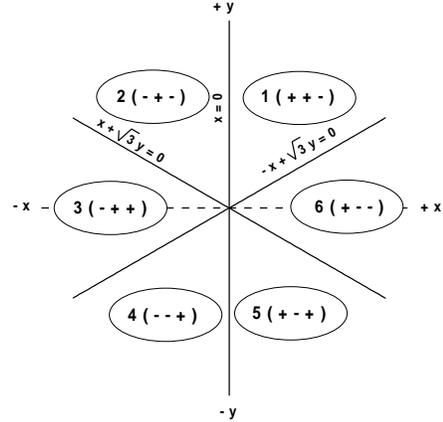}{0.9}    
\vskip -1cm
\caption{Six regions 1,2, $\cdots$, 6  of the
three-particle configuration that are divided by three
lines $x_{12}=0,\; x_{23} =0$ and $x_{31}=0$.
In this connection, see (\ref{15}) and (\ref{16}).
The order of entries in $(++-)$ etc. refer to the signs of
$x_{12}, \; x_{23}$ and $x_{31}$, respectively.
}
\label{fig2}
\end{figure}

\noindent
where $\kappa >0$ is a constant that is to be determined. 
The $\phi (x,y)$ is totally symmetric with respect to the 
interchange of any pair of particles 1, 2 and 3. Let us 
assume that the wave function of the bound state in each 
of the regions is of the form of 
\begin{equation}
\psi _{\nu} (x,y) = C_{\nu} \phi (x,y)\, , \;\;\;
\nu =1,2, \cdots , 6\, .
\label{21}
\end{equation}
The energy $E$ of the bound state is given by
\begin{equation}
E= - \frac{2\kappa^2}{m}\, .
\label{22}
\end{equation}

In order to satisfy the Schr\"{o}dinger equation in the 
entire space, the wave function has to satisfy the boundary 
conditions at $x=0$, etc. where the point interactions act. 
This can be done as follows. Let us start with region 1 by 
assuming the wave function $\psi_1 (x,y) = C_1 \phi (x,y)$
and apply the boundary condition (\ref{1}) to determine
the wave function in region 2. In doing so, we do not have
to be concerned with the $y$-dependence of the wave 
function. In the vicinity of the line $x=0$, the 
Schr\"{o}dinger equation (\ref{19}) can essentially be 
reduced to the two-particle equation (\ref{16}). Note that
in going from region 1 to region 2, $x$ changes from 
positive to negative. 
See, the signs of 
$x_{12}=(x_1 -x_2)/\sqrt{2}$, $x_{23}$ and $x_{31}$ shown 
in figure 2. We thus obtain
\begin{equation}
C_2 = \frac{C_1}{\eta}, 
\label{23}
\end{equation}
where $\eta$ is given by (\ref{11}). The $\kappa$ that 
appears in $\eta$ is that of (\ref{9}), the same $\kappa$
as that of the two-body case.

Next let us turn to the relation between $C_2$ and $C_3$.
It is convenient to introduce another set of Jacobi 
coordinates $x'$, $y'$ and $z'$, which are respectively
defined in terms of $x$, $y$ and $z$ of (\ref{17}) in 
which $x_1$, $x_2$ and $x_3$ are replaced by $x_3$, $x_1$ 
and $x_2$. Then the line that separates regions 2 and 3 
is represented by $x'=0$. We start in region 2 with 
the wave function $C_2 \phi (x',y')$. Recall that 
$\phi(x,y)=\phi(x',y')$, which is totally symmetric as 
we stated below (\ref{20}). The boundary condition along 
$x'=0$ leads to the wave function $C_3 \phi(x',y')$ of 
region 3 where $C_3 = \eta C_2 = C_1$. The reason why we 
obtain the factor $\eta$ rather than $1/\eta$ as in 
(\ref{23}) is that, in going from region 2 to region 3, 
$x'$ changes from negative to positive. Repeating 
similar steps we arrive at 
\begin{equation}
C_1 = C_3 = C_5\, , \hspace{0.4 cm} 
C_2 = C_4 = C_6 = \frac {C_1}{\eta}\, .
\label{24}
\end{equation}
If there are two possible values of $\kappa$, there are two 
bound states of the three-body system. For the coefficients
$C_{\nu}$'s of the two bound states, a relation of the form 
of (\ref{13}) holds for any two adjacent regions like 
1 and 2. This leads to the orthogonality between the two 
states.

The $\kappa$ and hence the energy of the bound state is
independent of $\theta$. The probability distribution
$|\psi_\nu (x,y)|^2$ is also independent of $\theta$.
The wave function and the energy are both smooth functions 
of the parameters of the interaction. Start with arbitrary 
values of the parameters. Let them continuously vary and 
approach the values of (\ref{4}), then we obtain McGuire's
results for the $\delta$-function potential. Let us assume
that no level crossing takes place in this limiting
procedure. In the limit of the $\delta$-function potential
we know that there can be only one bound state. It then 
follows that the bound state with the lower energy that we 
have obtained above is the ground state. 

If $\alpha=\gamma$ and $e^{i\theta}=1$, then $\eta =1$. 
The ground state is totally symmetric with respect to 
interchanges of the particles. The excited state, if it 
exists as we discussed below (\ref{11}), is totally 
antisymmetric. If $\alpha=\gamma$, and $e^{i\theta}=-1$, 
the symmetry and antisymmetry of the two states are reversed. 

Next let us consider the $N$-particle system. There are $N$! 
linear configurations of the $N$ particles. Assume that
the wave function is of the form of
\begin{equation}
\psi _\nu 
= C_\nu \exp(-\kappa \sum_{i>j} |x_{ij}|) \, ,
\;\;\; x_{ij} = \frac{1}{\sqrt{2}}(x_i - x_j)\, ,
\label{25}
\end{equation}
where $\nu$ refers to one of the $N$! configurations and 
$C_\nu$ is a constant coefficient associated with configuration 
$\nu$. It is understood that the centre-of-mass coordinate 
has been separated. Start with the configuration (1,2,3, 
$\cdots$ , $N$), for example, and proceed to (2,1,3, $\cdots$, 
$N$). Along the boundary between these two configurations, the
$N$-particle Schr\"{o}dinger equation is again reduced to the 
two-particle equation (\ref{16}). Therefore the coefficient
$C_2$ can be related to $C_1$ exactly in the same way as in
(\ref{23}), with the $\kappa$ of (\ref{9}).  When we go from 
one configuration to another, factor $\eta$ or $1/\eta$ enters 
for every permutation. The configurations can be grouped into 
those of even and odd permutations. The $C_\nu$'s are equal 
within each of the two groups. The $C_\nu$'s of different 
groups differ by factor $\eta$. Starting with the initial 
configuration, another configuration can be reached in 
different ways, but this does not give rise to any ambiguity 
in determining $C_\nu$'s. For the nomalization of wave 
functions of the form of (\ref{25}), see \cite{11}.

The energy of the bound state is given by
\begin{equation}
E = - \frac{\kappa ^2}{12m}N(N^2-1)\, ,
\label{26}
\end{equation}
where $\kappa$ is again that of (\ref{9}). There can be two
bound states. The $N$-dependence of $E$ is the same as that 
for the $\delta$-\-function potential. For the derivation of 
the $N$-dependent factor, see the appendix of McGuire's paper 
\cite{1}.

\section{Scattering states}
McGuire showed how the scattering or the transmission and
reflection problem for many-particle systems can be solved
exactly for the $\delta$-function potential \cite{1}. 
In this section let us examine the three-particle case. 
If the three-particle case can be solved, many-body cases
can be done in a similar manner as shown by McGuire.
The three-particle system can be 
regarded as one particle in two dimensions, as can be seen
from (\ref{19}). A wave propagates in the $x$-$y$ plane
and meets the interactions along the three lines $x=0$, etc. 
For the solvability of the problem \`a la McGuire, it is 
crucial that there is no diffraction. As McGuire showed, 
indeed there is no diffraction when the interaction is the 
$\delta$-function potential. For the general point 
interactions, however, there is diffraction.  This means 
that unfortunately McGuire's method as such does not work. 
This is what we are going to show below.

We will not review McGuire's calculation. Rather we simply 
apply it to the present case. Consider a ray (or a wave) 
incident in region 2 and transmitted and reflected by the 
potential barriers. Let us focus on the ray that goes out 
into region 1. There are two geometries as shown in 
figures 3 and 4. 

Assume that the amplitude of the incident wave is unity. 
In figure 3, the ray that goes through points A, B and 
A$^\prime$ obtains amplitude $T_{1-} R_{2+} R_{3+}$
and the other route of A, C and A$^\prime$ leads to  
$R_{1-} R_{2-} T_{3-}$ .
In figure 4, the amplitude of the outgoing ray is 
$R_{3-} T_{2+} R_{1+}$.
Here the $T$'s and $R$'s are the transmission and reflection 
coefficients that have been worked out before \cite{6,7};
we give them explicitly below. The suffices $+$ and $-$ 
indicate the direction of incidence, which were respectively 
denoted with R and L before. Suffices 1, 2 and 3 refer to 
the three angles of incidence,
\begin{figure}
\postscript{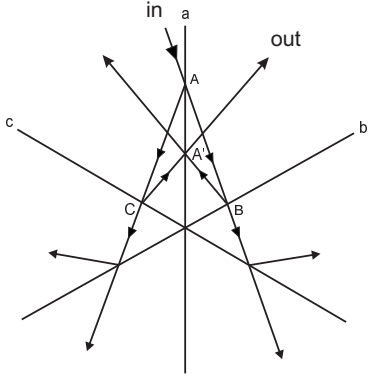}{0.8}    
\vskip -1.5cm
\caption{Ray diagram that applies when the incoming
hits the potential line $x=0\;$ i.e. $x_{12}=0$ first.
We focus on the ray that is incident in region 2 and
goes out in region 1.
For the regions, see figure 2.
}
\label{fig3}
\end{figure}
\vskip -0.5cm

\begin{figure}
\postscript{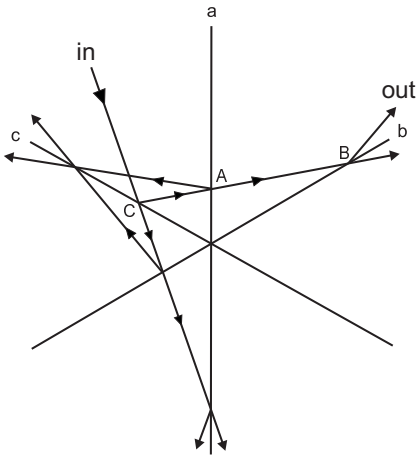}{0.9}    
\vskip -1.5cm
\caption{Ray diagram that applies when the incoming 
hits the potential line $(x+ \sqrt 3 y)=0\;$ i.e.
$x_{31}=0$ first. We focus on the ray that is incident
in region 2 and goes out in region 1.
For the regions, see figure 2.
}
\label{fig4}
\end{figure}

\noindent
\begin{equation}
\varphi _1 = \varphi\, , \;\;\;
\varphi _2 = \varphi + \frac{\pi}{3}\, , \;\;\;
\varphi _3 = -\varphi + \frac{\pi}{3}\, , \;\;\;
\label{27}
\end{equation}
which were defined by McGuire \cite{1}. The $T_{i\pm}$ and 
$R_{i\pm}$ are associated with the normal components of
$k$,
\begin{equation}
k_i = k \sin \varphi _i \, , \;\;\; i=1,2,3\, .
\label{28}
\end{equation}
Note that
\begin{equation}
k_1 + k_2 = k_3\, .
\label{29}
\end{equation}  

The path lengths of the rays in the two geometries are equal. 
If the two amplitudes associated with the outgoing rays that 
are shown in figures 3 and 4, i.e.,
\begin{equation}
R_{1-} R_{2-} T_{3-} + T_{1-} R_{2-} R_{3+} 
\;\;\; {\rm and}\;\;\; R_{3-} T_{2+} R_{1+} \; ,
\label{30}
\end{equation}
are equal, there is no diffraction. Then the wave function 
of the scattering state can be written down as was done by 
McGuire. This was indeed the case for the $\delta$-\-function 
potential. In that case the relation (\ref{29}) is instrumental. 

For the point interactions of the four-parameter family, the 
$T$'s and $R$'s read as \cite{6,7},
\begin{equation}
T_{\pm} = \frac{4i e^{\mp i\theta} km }{D}\, ,
\label{31}
\end{equation}
\begin{equation}
R_{\pm} = \frac{\delta k^2 \mp 2ikm(\alpha -\gamma) 
+4\beta m^2}{D}\, ,
\label{32}
\end{equation}
\begin{equation}
D = \delta k^2 +2ikm(\alpha +\gamma) -4\beta m^2\, .
\label{33}
\end{equation}
Note that $T_+ \neq T_-$ unless $e^{i\theta}$ is real.
The $R_{\pm}$ are independent of $\theta$. Although somewhat 
tedious it is straightforward to show that the condition for 
vanishing diffraction can be satisfied for arbitrary $k$ and 
$\varphi$ if and only if 
\begin{equation}
\alpha = \gamma\, ,\;\;\; \delta = 0\, ,
\;\;\; e^{i\theta} = \pm 1\, .
\label{34}
\end{equation}
If we combine (\ref{34}) with the constraint 
$\alpha \gamma -\beta \delta =1$, we obtain 
$\alpha=\gamma= \pm 1$. 
It is interesting that ADF \cite{7} arrived at exactly the same 
condition by examining the Yang-Baxter equation for the system
within the context of the Bethe ansatz \cite{2,12}.

If $e^{i\theta}=-1$, the interaction of $\alpha=\gamma= -1$ 
is the usual $\delta$-function potential of
(\ref{5}) and we obtain McGuire's solution. The other 
interaction of $\alpha=\gamma= 1$, which ADF called the
``anti-$\delta$ interaction", differs from the 
$\delta$-function potential only through the different 
choice of $e^{i\theta}$. 
For these two interactions, 
the wave functions of the three-body system 
only differ in the sign that depends on the six regions;
the scattering amplitudes of (\ref{30}) 
differ only through their overall sign. 
For the physical quantities of the system, the two
interactions end up with identical results. 
In this sense they are essentially equivalent. 
See, however, appendix 1. 

\section{Summary and Discussion}
We have attempted to extend McGuire's work \cite{1} on the 
many-particle system interacting through the 
$\delta$-function potential to accommodate the four-parameter 
family of point interactions. We have succeeded in doing 
so for the bound states, but not quite for the scattering 
states. For the bound states we obtained exact solutions. 
If the interaction supports one or two bound states for 
the two-body system, it supports the same number of bound
states for an $N$-body system. It is interesting that, 
unlike McGuire's case, there can be two bound states (for 
the same system with the same interaction). We suspect that
what we have found exhausts all possible bound states, but
we have no rigorous proof for that.

The scattering problem becomes complicated in general 
because of the emergence of diffraction. McGuire's method
as such does not work unless the parameters are restricted
to (\ref{34}). There is no such restriction in the bound 
state problem. The condition of no diffraction is 
instrumental for the scattering problem but it is not for 
the bound state. We admit that we were surprised by this 
finding. The emergence of diffraction does not necessarily 
mean the nonexistence of scattering solutions. We believe 
that scattering solutions exist even in the presence of 
diffraction. The solutions would require more sophisticated 
approach, probably, such as those developed by Albeverio, 
McGuire and Hurst, and Lipszyc \cite{13}. In the cases 
that were examined in \cite{13} diffraction occurs
because the particles have different masses, or the 
interactions for different pairs are different, and so on.
Still scattering solutions can be constructed.

ADF \cite{7} independently examined essentially the 
same problem as what we have done except that they also
considered the impenetrable type of point interactions
that we have not considered. Let us comment only on their
results on the penetrable type. 
For the scattering problem, ADF's results and ours agree. 
For the bound state problem, ADF obtained solutions only 
for the parameter sets for which they found scattering 
solutions. As we emphasized already, the bound solutions 
that we found have no such restrictions. 
ADF's solutions are special cases of our solutions. 
As we said above there can be two bound states. 
ADF's solutions have no such possibility.

\section*{Acknowledgements}
One of the authors (YN) is grateful to Universidade de S\~ao 
Paulo and Instituto de F\'\i sica Te\'orica of Universidade 
Estadual Paulista for warm hospitality extended to him 
during his visit of 1998. This work was partially supported 
by Funda\c c\~ao de Amparo \`a Pesquisa do Estado de S\~ao 
Paulo (FAPESP), Conselho Nacional de Desenvolvimento 
Cient\'\i fico e Tecnol\'ogico (CNPq) and the Natural 
Sciences and Engineering Research Council of Canada.

\vskip 1cm
\newpage

\begin{center}
{\bf Appendix 1: 
Symmetry of the bound state wave functions}
\end{center} 

In this appendix we assume $e^{i\theta}=\pm 1$.
Let us consider the bound states of the one-particle 
system that we discussed in section 2, in particular, 
the case of $\alpha=\gamma$ and $\delta > 0$. Since 
$\alpha = \gamma$, the system is invariant under space 
reflection, and parity is a good quantum number. Assume
$\alpha <1$ so that there are two bound states. 

As we pointed out in section 2, if $e^{i\theta}= 1$ 
the parity is even for the ground state and is odd for 
the excited state. On the other hand, if 
$e^{i\theta}=-1$ the parity is odd for the ground 
state and is even for the excited state. This may sound 
odd but there is nothing wrong with this in principle.

Let us consider an $N$-body system with the same 
interactions assumed above. If there are two bound states 
in the two-body case, there are also two bound states
in the $N$-body case. If $e^{i\theta}=1$, the 
ground state is totally symmetric and the excited state 
totally antisymmetric. If $e^{i\theta}=-1$, the 
symmetry and antisymmetry of the two states are reversed.
A totally symmetric (antisymmetric) state can accommodate 
bosons (fermions). There is ``duality" between the 
boson systems. All physically observable quantities
are the same between the two systems.

Let us add that, with $e^{i\theta}=1$, the wave function 
of a bound state for an attractive $\delta '$ interaction 
is of the same form as that of an attractive 
$\delta$-function potential. This may seem to imply a kind
of duality. The $\delta '$ and $\delta$ interactions, 
however, give different transmission and reflection 
coefficients. In this sense this duality is restricted to 
bound states.
\vskip 1cm
\newpage

\begin{center}
{\bf Appendix 2: The $\delta$-function potential}
\end{center} 

In order to facilitate comparison between our results and
those of McGuire \cite{1}, let us summarize the case in 
which we start with the two-particle interaction
\begin{equation}
V(x_i -x_j) = g_0 \delta (x_i -x_j)\, .
\label{b1}
\end{equation}
McGuire calls $g_0$ (which he denotes with $C$) the true 
strength of the $\delta-$\-function interaction. Then, as we
said below (\ref{16}), the $V(x)$ of (\ref{16}) is given by
\begin{equation}
V(x) = g\delta (x)\, ,\;\;\; g =\frac{g_0}{\sqrt 2}\, .
\label{b2}
\end{equation}
If $g<0$, there is one bound state in each of the 
two-particle and many-particle systems with
\begin{equation}
\kappa = -gm = -\frac{g_0 m}{\sqrt 2}\, .
\label{b3}
\end{equation}
The energy of the $N$ particle bound state is
\begin{equation}
E = - \frac{g^2 m}{12} N(N^2 -1)
= - \frac{{g_0}^2 m}{24} N(N^2 -1)\, .
\label{b4}
\end{equation}

If we put $m=1$ and $g_0 =C=-g_{MG}/\sqrt{2}$, where $g_{MG}$
is McGuire's $g$, the $E$ of (\ref{b4}) becomes the same
as McGuire's $E$ (p. 634 of \cite{1}). There is a misprint
in McGuire's three-body wave function (p. 630 of \cite{1}).
If we put $m=1/2$ and $g_0 = - g_{CD}$, where $g_{CD}$ is the 
$g$ of \cite{11}, (\ref{b4}) is reduced to (9) of \cite{11}.



\begin{references}
\bibitem{1} McGuire J B 1964 J. Math. Phys. {\bf 5} 622;
1965 ibid {\bf 6} 432
\bibitem{2} Yang C N 1967 Phys. Rev. Lett. {\bf 19} 1312;
1968 Phys. Rev. {\bf 168} 1920
\bibitem{3} Albeverio S, Gesztesy F, H$\phi$egh-Krohn R and 
Holden H 
1988 {\it Solvable Models in Quantum Mechanics} 
(Berlin: Springer)
\bibitem{4} \v{S}eba P 1986 Czech. J. Phys. B {\bf 36} 667; 
1986 Rep. Math. Phys. {\bf 24} 111
\bibitem{5} Gesztesy F and Holden H 1987 J. Phys. A: Math. 
Gen. {\bf 20} 5157
\bibitem{6} Chernoff P R and Hughes R J 1993
J. Functional Analysis {\bf 111} 97
\bibitem{7} Coutinho F A B, Nogami Y and Perez J F 1997 
J. Phys. A: Math. Gen. {\bf 30} 3937 (1997) 
and earlier references quoted there
\bibitem{8} Coutinho F A B, Nogami Y and Perez J F 1998
{\it Time-reversal aspect of the point interactions in 
one-dimensional quantum mechanics}  
J. Phys. A: Math. Gen. in press
\bibitem{9} Albeverio S, D\c{a}browski L and Fei S-M 1998
{\it One dimensional many-body problems with point 
interactions} SISSA 139/FM/98
\bibitem{10} Albeverio S, D\c{a}browski L and Kurasov P 1998
Lett. Math. Phys. {\bf 45} 33
\bibitem{11} Calogero F and Degasperis A 1975 Phys. Rev. A
{\bf 11} 265
\bibitem{12} Baxter R J 1972 Ann. Phys. {\bf 70} 193;
1978 Phil. Trans. Royal Soc. London {\bf 289} 315
\bibitem{13} Albeverio S 1969 Helv. Phys. Acta {\bf 40} 135; \\
McGuire J B and Hurst C A 1972 J. Math. Phys. {\bf 13} 1595; \\
Lipszyc K 1972 Acta Phys. Pol. A {\bf 42} 571;
1974 J. Math. Phys. {\bf 15} 133;
1980 J. Math. Phys. {\bf 21} 1092
\end{references}
\end{document}